\documentclass[a4paper,fleqn,usenatbib,useAMS]{mnras}

\usepackage{graphicx}	
\usepackage{amsmath}	
\usepackage{amssymb}	
\usepackage{multicol}        
\usepackage{bm}		
\usepackage{pdflscape}	
\usepackage{array}

\usepackage[T1]{fontenc}
\usepackage{ae,aecompl}
\usepackage{natbib} 

\usepackage{float}

\title[Inner Accretion Disc Collapse Timescales]{X-ray Dips in AGN and Microquasars -- Collapse Timescales of Inner Accretion Disc}

\author[Shende et al.]{{Mayur B. Shende$^{1}$\thanks{E-mail: \href{mailto:mayur.shende@students.iiserpune.ac.in}{mayur.shende@students.iiserpune.ac.in}}}, {Prashali Chauhan$^{2}$\thanks{E-mail:  \href{prchauha@syr.edu}{prchauha@syr.edu}}} and {Prasad Subramanian$^{1}$\thanks{E-mail: \href{mailto:p.subramanian@iiserpune.ac.in}{p.subramanian@iiserpune.ac.in}}}
\\
$^{1}$Indian Institute of Science Education and Research, 
Dr. Homi Bhabha Road, Pashan, Pune 411008,
India\\
$^{2}$Department of Physics, Syracuse University, USA}

\pubyear{2019}

\begin{document}
\label{firstpage}
\pagerange{\pageref{firstpage}--\pageref{lastpage}}
\maketitle

\begin{abstract}
The temporal behaviour of X-rays from some AGN and microquasars is thought to arise from the rapid collapse of the hot, inner parts of their accretion discs. The collapse can occur over the radial infall timescale of the inner accretion disc. However, estimates of this timescale are hindered by a lack of knowledge of the operative viscosity in the collisionless plasma comprising the inner disc.  
We use published simulation results for cosmic ray diffusion through turbulent magnetic fields to arrive at a viscosity prescription appropriate to hot accretion discs. We construct simplified disc models using this viscosity prescription and estimate disc collapse timescales for 3C 120, 3C 111, and GRS 1915+105. The Shakura-Sunyaev $\alpha$ parameter resulting from our model ranges from 0.02 to 0.08. Our inner disc collapse timescale estimates agree well with those of the observed X-ray dips. We find that the collapse timescale is most sensitive to the outer radius of the hot accretion disc. 
\end{abstract}





\section{Introduction} \label{intro}

It is well established by now that active galactic nuclei (AGN) and galactic microquasars are powered primarily via accretion of gas onto the black hole at their centre. Among the various interesting aspects of the accretion phenomena, one that stands out is the often complex behaviour of X-ray intensity from such sources. X-rays from these sources are commonly believed to emanate from the hot inner parts of the accretion discs surrounding the central black hole. Consequently, the interpretations of the observed temporal behaviour of X-rays rely on the dynamics of these hot accretion discs. The well studied galactic microquasar GRS 1915+105 exhibits X-ray behaviour that is attributed to the rapid removal and gradual replenishment of the inner accretion disc (e.g., \citealt{belloni1997}). Subsequent studies of this source \citep{yadav1999,naik2001,Vadawale2001,Vadawale2003} base their interpretations on the dynamics of the sub-Keplerian, post-shock halo of a two-component advective flow around the central black hole. On the other hand, X-ray dips are reported from the AGN sources 3C120 and 3C 111 \citep{chatterjee2009,Chatterjee2011,marscher2002,marscher2006}. Interestingly, such X-ray dips are thought to be related to episodic ejections in these sources, forming a blobby jet. Such episodic plasma ejections, which could be launched in a manner similar to solar coronal mass ejections (as envisaged by \citealt{shende2019}), have also been reported from GRS 1915+105 by \cite{naik2001}, \cite{Vadawale2001} and \cite{Vadawale2003}.

In this work, we concentrate on the interpretation of the infall/collapse timescales of the hot, inner accretion disc in such sources. For GRS 1915+105, this would correspond to the fast rise timescales of a few seconds in the outburst state, which has been attributed to the infall timescale of the inner disc (the quantity $t_{\rm vis}^{\rm h}$ in \citealt{yadav1999}). We show that this infall timescale corresponds to the viscous timescale in the hot inner disc. We emphasize that this is different from the slow viscous timescale of the cold, outer disc that is held responsible for the regeneration of the entire inner disc in \cite{belloni1997}. For the extragalactic sources 3C 120 and 3C 111, we seek to interpret the X-ray dip timescales. The main feature of our work here lies in employing a physical viscosity mechanism (instead of relying on values assigned to the Shakura-Sunyaev $\alpha$ parameter) to estimate the viscous timescale in the hot inner disc. We consider hot, two-temperature accretion discs, where the electron temperatures are $\approx 10^{9}$ K and the proton temperatures are in excess of $10^{11}$ K. The hot protons in the two-temperature plasma are effectively collisionless, much like cosmic rays. We use published cosmic ray diffusion coefficients in turbulent magnetic fields to estimate the effective viscosity in hot accretion discs and construct simplified disc models using it. The infall timescale is calculated using these disc models.

\section{Hybrid viscosity in hot accretion discs}

The interpretation for most of the observations referred above invoke a hot inner accretion disc/corona which is responsible for Comptonized X-rays. Our aim in this paper is to identify the collapse timescale of the inner accretion disc with the viscous infall timescale in this region. Starting with early work (e.g., \citealt{kafatos1988,belloni1997}) and continuing onward, researchers have typically parametrized the viscous timescale in terms of the dimensionless $\alpha$ parameter \citep{shakura1973}. The values of $\alpha$ are not very well constrained, and this reflects the lack of our knowledge concerning the microphysical nature of viscosity. This question is especially pronounced in the hot inner regions of accretion discs, where the plasma is so hot that it is collisionless; i.e., the mean free path of hot protons typically exceeds macroscopic lengths such as the disc height. The operative viscosity in such a situation cannot be due to proton-proton collisions, since they are very rare. On the other hand, \cite{sbk96} (SBK96 from now on) showed that magnetic irregularities arising out of turbulence can act as scattering centres for protons. These proton-magnetic scattering centre collisions (essentially interactions between protons and the turbulent wave spectrum) could give rise to a ``hybrid'' kind of viscosity, that is neither due to particle collisions alone nor due to Reynolds stresses from the magnetic field (alone). The work of SBK96 was limited to considering only a single lengthscale in the magnetic irregularity spectrum. Furthermore, it did not take into account the likely presence of a large scale, ordered toroidal magnetic field embedded in the accretion disc - a feature suggested by simulations of the magnetorotational instability \citep{balbus1998,Quataert2002,sharma2003} which is thought to be responsible for the generation of magnetic turbulence. On the other hand, studies of cosmic ray propagation and scattering in the presence of turbulent magnetic irregularities have a long history (e.g., \citealt{parker1965,Jokipii1966,Giacalone1999}). This is a problem similar to ours, in that the energetic cosmic ray protons are collisionless, and are scattered by irregularities arising out of magnetic turbulence. The diffusion coefficient obtained for this situation can be used to determine the operative mean free path (and thereafter, the relevant viscosity coefficient) in hot accretion discs. Some simulation studies of cosmic ray transport give convenient analytical fits to their results \citep{casse2002,candia2004,snodin2016}.

In this work, we use the analytical fits to diffusion coefficients provided by \cite{candia2004} and \cite{snodin2016} to compute the hybrid viscosity in the accretion discs arising due to the hot protons diffusing through tangled magnetic fields. Their results take into account the presence of a large scale ordered field and a small scale, turbulent field. They give expressions for all the components of the diffusion tensor for cosmic rays - this includes the diffusion coefficient for transport parallel to the large scale field ($D_{||}$) and perpendicular to it ($D_{\perp}$). We extract expressions for the effective mean free paths from the diffusion coefficients and use it to compute the viscosity relevant to our situation. 
As mentioned earlier, there is evidence for a large scale toroidal magnetic field embedded in magnetized accretion discs (e.g., \citealt{matsumoto1995}). Since viscous angular momentum transport in accretion discs is primarily responsible for radially inward accretion, the main focus is on the $r-\phi$ component of the viscosity tensor \citep{shakura1973,frank2002}. Since there is a large-scale toroidal ($\hat{\phi}$ directed) field in the accretion disc, we are interested in the mean free path derived from $D_{\perp}$ (since the radial direction is $\perp {\hat{\phi}}$ ) to compute the effective viscosity. 

\subsection{Hybrid Viscosity using $D_{\perp}$ from Candia \& Roulet} \label{visc}

In what follows, we use the expression for perpendicular diffusion coefficient $D_{\perp}$ provided by \cite{candia2004}. A preliminary account of this treatment was given in \cite{subramanian2005}.

From Eqs. (18) and (19) of \cite{candia2004}, we get the expression for $D_{\perp}$ as
\begin{equation}
D_{\perp} = v_{\rm rms} H D_{c}  \,\,\,\,\,\,\,\,\,\,\,\,  \rm cm^2 s^{-1} \label{eq:a}
\end{equation}

The disc height $H$ is taken to be the representative macroscopic scale length and the proton rms speed $v_{\rm rms} \equiv \sqrt{3 k_{\rm B} T_{i} / m_{\rm p}}$ (where $k_{\rm B}$, $T_{i}$ and $m_{\rm p}$ denote Boltzmann's constant, proton temperature and proton mass respectively) is used as a representative speed. Hot ($\sim 10^{12}$K) protons we consider have energies of the order of $86$ MeV, and are therefore non-relativistic.
The quantity $D_{c}$ is given by the expression
\begin{equation}
\begin{split}
D_{c} = N_{\perp} (\sigma^2)^{a_{\perp}} \frac{N_{\parallel}}{\sigma^2} \bigg [ \bigg (\frac{\rho}{\rho_{\parallel}} \bigg )^{2(1-\gamma)} + \bigg (\frac{\rho}{\rho_{\parallel}} \bigg )^{2} \bigg ]^{1/2} \\ \times \left\{
\begin{array}{ll}
\rho & \quad 0 < \rho \leq 0.2 \\
0.04/\rho & \quad 0.2 < \rho < 1
\end{array}
\right.     \label{eq:b}
\end{split}
\end{equation}
where $\sigma^2 \equiv \langle B_{r}^{2} \rangle /\langle B_{0}^{2} \rangle $ is a measure of the turbulence level and defined by the ratio of energy density in the small scale turbulent magnetic fields ($B_{r}^{2}$) to that in the large-scale field ($B_{0}^{2}$). The quantity $\rho \equiv r_{L}/H$ represents magnetic rigidity and is defined as the ratio of the Larmor radius of a proton ($r_{L}$) to a macroscopic scale length (which we take to be the disc height $H$). The parameters $a_{\perp}$, $N_{\perp}$, $N_{\parallel}$, $\rho_{\parallel}$, and $\gamma$ are specific to different kinds of turbulence, and are defined in Table 1 of \cite{candia2004}.

The coefficient of dynamic viscosity is usually defined as 
\begin{equation}
\eta \,\,({\rm g \,\,cm^{-1} s^{-1}}) = N m v \lambda\, , 
\label{eq:b1}
\end{equation}
where $N$ and $m$ are the number density and mass of the relevant particles, $v$ is the relevant velocity (in our case, the thermal rms velocity $v_{\rm rms}$) and $\lambda$ is the relevant mean free path \citep{spitzer1962,mihalas1984}. On the other hand, the diffusion coefficient $D$ is usually defined as 
\begin{equation}
D \,\, ({\rm cm^{2} \,\, s^{-1}}) = v \lambda\, , 
\label{eq:b2}
\end{equation}
where $v$ is the relevant velocity and $\lambda$ is the relevant mean free path. Combining Eqs.~(\ref{eq:b1}) and (\ref{eq:b2}), we get $\eta = N m D$. The coefficient of dynamic viscosity relevant for our situation is therefore
\begin{equation}
\eta_{\rm hyb} = N_{i} m_{\rm p} D_{\perp} ,    \label{eq:c}
\end{equation}

\subsection{Solutions for disc model} \label{dm}

The basic equations for the structure of the hot, inner part of accretion disc are given in Appendix \ref{append}. These equations, which approximate the relevant differential equations by scaling relations (e.g., $\partial/\partial R \rightarrow 1/R$, $\partial/\partial z \rightarrow 1/H$) are essentially the same as those for the two-temperature accretion disc solution originally put forth by \cite{shapiro1976} and used in \cite{eilek1983} and SBK96. Using Eq.~(\ref{eq:a}) and Eq.~(\ref{eq:ag}), Eq.~(\ref{eq:c}) can be written as
\begin{equation}
\eta_{\rm hyb} = 2.52 \,\, \tau_{\rm es} v_{\rm rms} D_{c}   \label{eq:d}
\end{equation}

Using the hybrid viscosity prescription of Eq.~(\ref{eq:d}) in Eq.~(\ref{eq:aa}), using Eqs.~(\ref{eq:ae}), (\ref{eq:ah}) and (\ref{eq:aj}), and assuming $T_{i} \gg T_{e}$, we get the following expression for the Shakura-Sunyaev viscosity parameter $\alpha_{\rm hyb}$:
\begin{equation}
\alpha_{\rm hyb} = 2.63 \,\, f_{1}^{-1/2} D_{c}     \label{eq:e}
\end{equation}

Using Eqs.~(\ref{eq:ad}), (\ref{eq:ag}), (\ref{eq:ah})--(\ref{eq:aj}) and Eq.~(\ref{eq:e}) for $\alpha_{\rm hyb}$, we get the following implicit equation for the electron scattering optical depth $\tau_{es}$:
\begin{equation}
\tau_{\rm es}^{3} (1 + \tau_{\rm es})^{3/2} = 272.5 \,\, (\ln{\Lambda})^{-1} \bigg( \frac{\dot{M}}{\dot{M}_{E}} \bigg)^{1/2} f_{1}^{-1/2} f_{2}^{-1/2} f_{3} y^{3/2} \alpha_{\rm hyb}^{1/2} R_{*}^{-3/4}          \label{eq:f}
\end{equation} 

The expressions for $\alpha_{\rm hyb}$(Eq.~\ref{eq:e}) and $\tau_{\rm es}$(Eq.~\ref{eq:f}), along with Eqs.~(\ref{eq:aa})--(\ref{eq:aj}) yield the following self-consistent solutions for the disc model:
\begin{equation}
T_{i} = 4.1 \times 10^{12} \bigg(\frac{\dot{M}}{\dot{M}_{E}}\bigg) f_{1}^{1/2} f_{2} \tau_{\rm es}^{-1} D_{c}^{-1} R_{*}^{-3/2}       \label{eq:g}
\end{equation} 

\begin{equation}
T_{e} = 1.48 \times 10^{9} y \tau_{\rm es}^{-1} [g(\tau_{\rm es})]^{-1}   \label{eq:h}
\end{equation}

\begin{equation}
N_{i} = 1.65 \times 10^{11} \bigg(\frac{\dot{M}}{\dot{M}_{E}} \bigg)^{-1/2} {M_8}^{-1} f_{1}^{1/4} f_{2}^{-1/2} D_{c}^{1/2} \tau_{\rm es}^{3/2} R_{*}^{-3/4}       \label{eq:i}
\end{equation}

\begin{equation}
\frac{H}{R} = 0.62 \,\, \bigg( \frac{\dot{M}}{\dot{M}_{E}} \bigg)^{1/2} f_{1}^{-1/4} f_{2}^{1/2} \tau_{\rm es}^{-1/2} D_{c}^{-1/2} R_{*}^{-1/4}    \label{eq:j}
\end{equation}

In these equations, the quantity $R_{*}$ represents the disc radius in units of the gravitational radius ($R_{\rm g} = GM/c^2$), $M_{8}$ represents the black holes mass in units of $10^{8}$ solar masses, $\dot{M}/\dot{M}_{E}$ represents the accretion rate in units of Eddington rate ($\dot{M}_{E} \equiv L_{\rm E}/c^2$, where $L_{\rm E} \equiv 4 \pi G M m_{\rm p} c /\sigma_{\rm T}$ is the Eddington luminosity and $\sigma_{\rm T}$ is the Thomson electron scattering cross section), $T_{i}$ is the proton temperature and $T_{e}$ is the electron temperature (in Kelvin), $N_{i}$ is the number density of protons in units of ${\rm cm}^{-3}$, and $y$ is the Compton y-parameter.

\subsection{Model Self-Consistency Conditions}   \label{sc}

Before evaluating the viscous timescales for disc models, we note that the following self-consistency conditions need to be satisfied by each of the models we consider:

\begin{enumerate}
\item $H/R \lesssim 1$ : the slim disc condition, which is implicit in the equation of vertical hydrostatic equilibrium (Eq.~\ref{eq:ab}). 
\item $T_{i} \gg T_{e}$ : the two-temperature condition.
\item $\lambda_{ii} / H \gg 1$ and $\lambda/H\, , \, \lambda/R < 1$ : the proton-proton mean free path $\lambda_{ii}$ is much larger than the disc height (i.e., the protons are collisionless), but the effective mean free path $\lambda$ (arising out of proton-turbulent wave spectrum interaction) is smaller than $H$.
 
\end{enumerate}
 
\begin{figure}
	\includegraphics[width=\columnwidth]{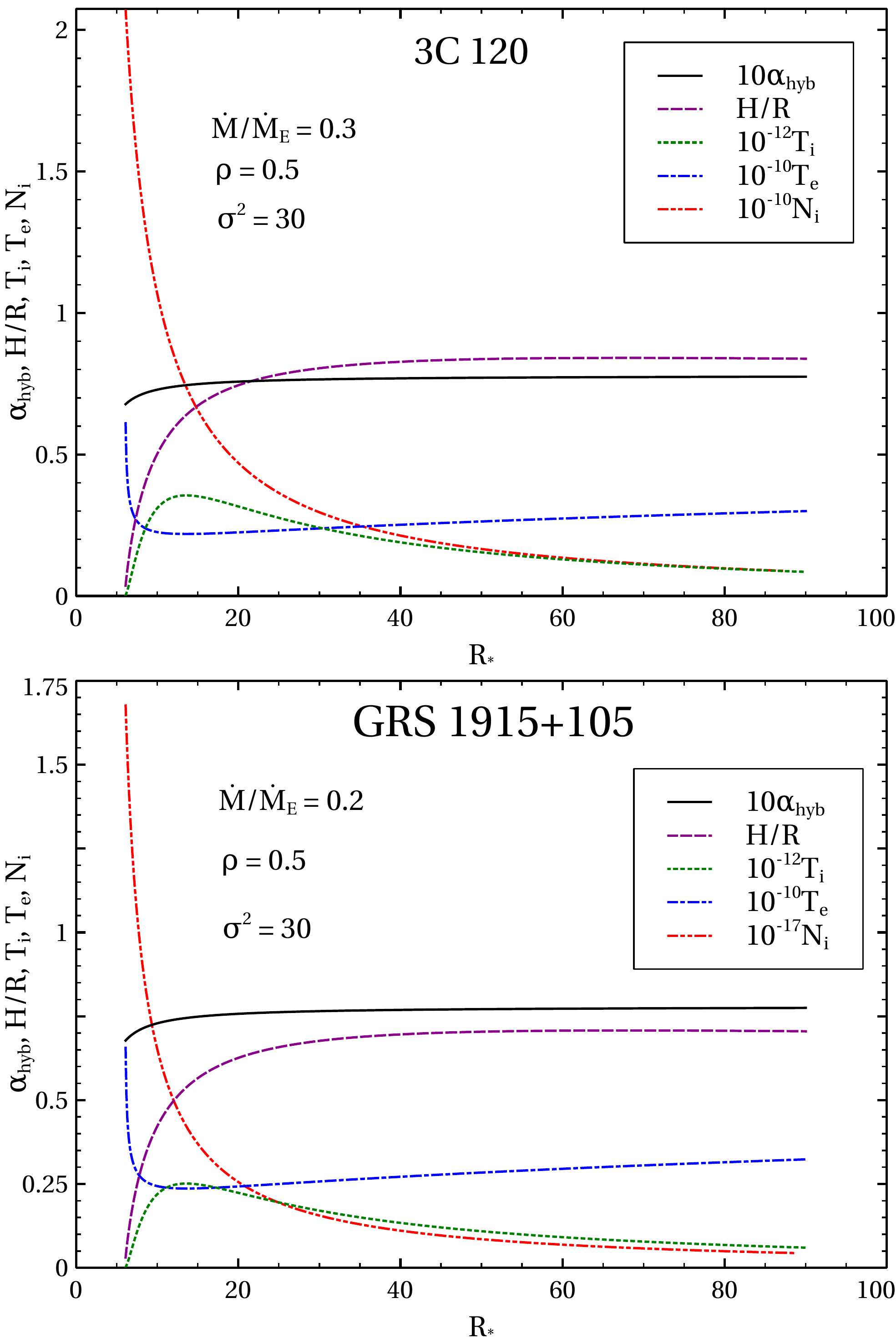}
	\caption{Representative accretion disc models for 3C120 (upper panel) and GRS 1915+105 (lower panel). This figure shows results for $10 \alpha_{\rm hyb}$, $H/R$, $10^{-12}T_{i}$, $10^{-10}T_{e}$ and $10^{-10}N_{i}$, with $\dot{M}/\dot{M}_{E}=0.3$, $\rho=0.5$ and $\sigma^{2}=30$.}
	\label{fig:diskpara}
\end{figure}

\begin{figure}
	\includegraphics[width=\columnwidth]{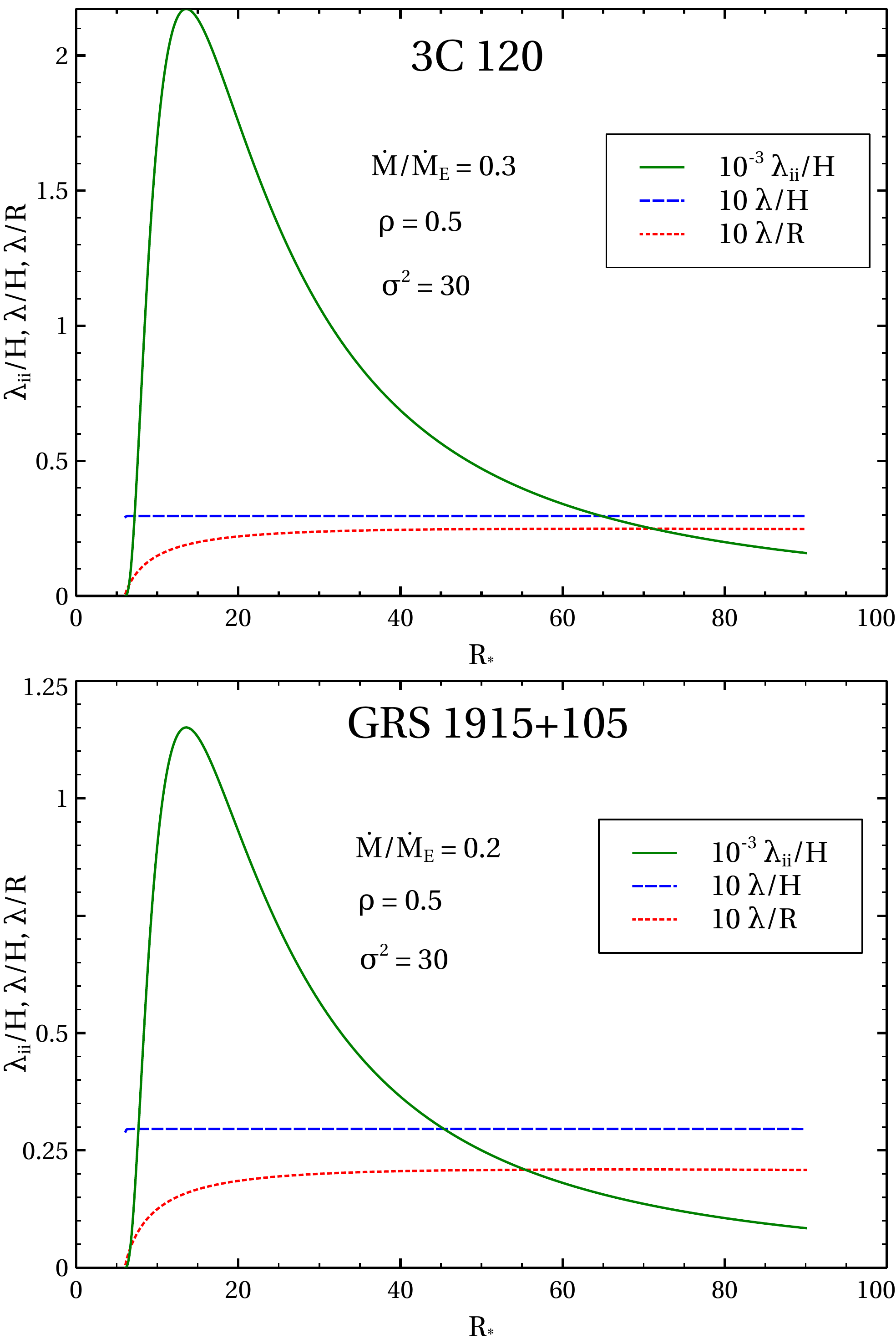}
	\caption{Representative accretion disc models for 3C120 (upper panel) and GRS 1915+105 (lower panel). This figure shows results for $10^{-3} \lambda_{ii}/H$, $10\lambda/H$ and $10 \lambda/R$, with $\dot{M}/\dot{M}_{E}=0.3$, $\rho=0.5$ and $\sigma^{2}=30$.}
	\label{fig:selfconsist}
\end{figure}

Figure~(\ref{fig:diskpara}) shows representative/fiducial accretion disc models for 3C 120 (upper panel), and GRS 1915+105 (lower panel). The parameters used are $a_{*} = 0$ (Swarzschild black hole), $y=1$, $\ln{\Lambda}=15$, $\rho=0.5$ and $\sigma^{2}=30$. The accretion rate is taken to be $\dot{M}/\dot{M}_{E}=0.3$ for 3C 120 \citep{chatterjee2009}, and $0.2$ for GRS 1915+105 \citep{zdziarski2016}. The black hole mass is taken to be $M_{8} = 0.55$ for 3C 120 \citep{peterson2004}, and $M_8 = 1.24 \times 10^{-7}$ for GRS 1915+105  \citep{reid2014}. The black solid lines show $10 \alpha_{\rm hyb}$, while the magenta dashed lines represent $H/R$. The green dotted lines show the proton temperatures in units of $10^{12} K$, while the electron temperatures in units of $10^{10}K$ are represented by blue dashed-dot lines. The proton number densities are depicted by red dashed-dot-dot lines, in units of $10^{10}\,{\rm cm^{-3}}$ for 3C 120, and in units of $10^{17}\,{\rm cm^{-3}}$ for GRS 1915+105. The self-consistency constraints discussed in \S~\ref{sc} are depicted in Figure~(\ref{fig:selfconsist}). The green solid lines show $10^{-3}\lambda_{ii}/H$, while the dashed blue lines represent $10\lambda/H$. The quantity $10\lambda/R$ is shown by the red dotted lines. None of the quantities, save for the proton number density ($N_{i}$), depend upon the black hole mass (Eqs~\ref{eq:e}, \ref{eq:g}--\ref{eq:j}). This is evident from Figure~(\ref{fig:diskpara}), which shows that the plasma is collisionless and two-temperature (with proton temperatures $T_{i} \approx 10^{11}$ K and electron temperatures $T_{e} \approx 10^{9}$ K) for 3C 120 as well as GRS 1915+105.

\begin{figure}
	\includegraphics[width=\columnwidth]{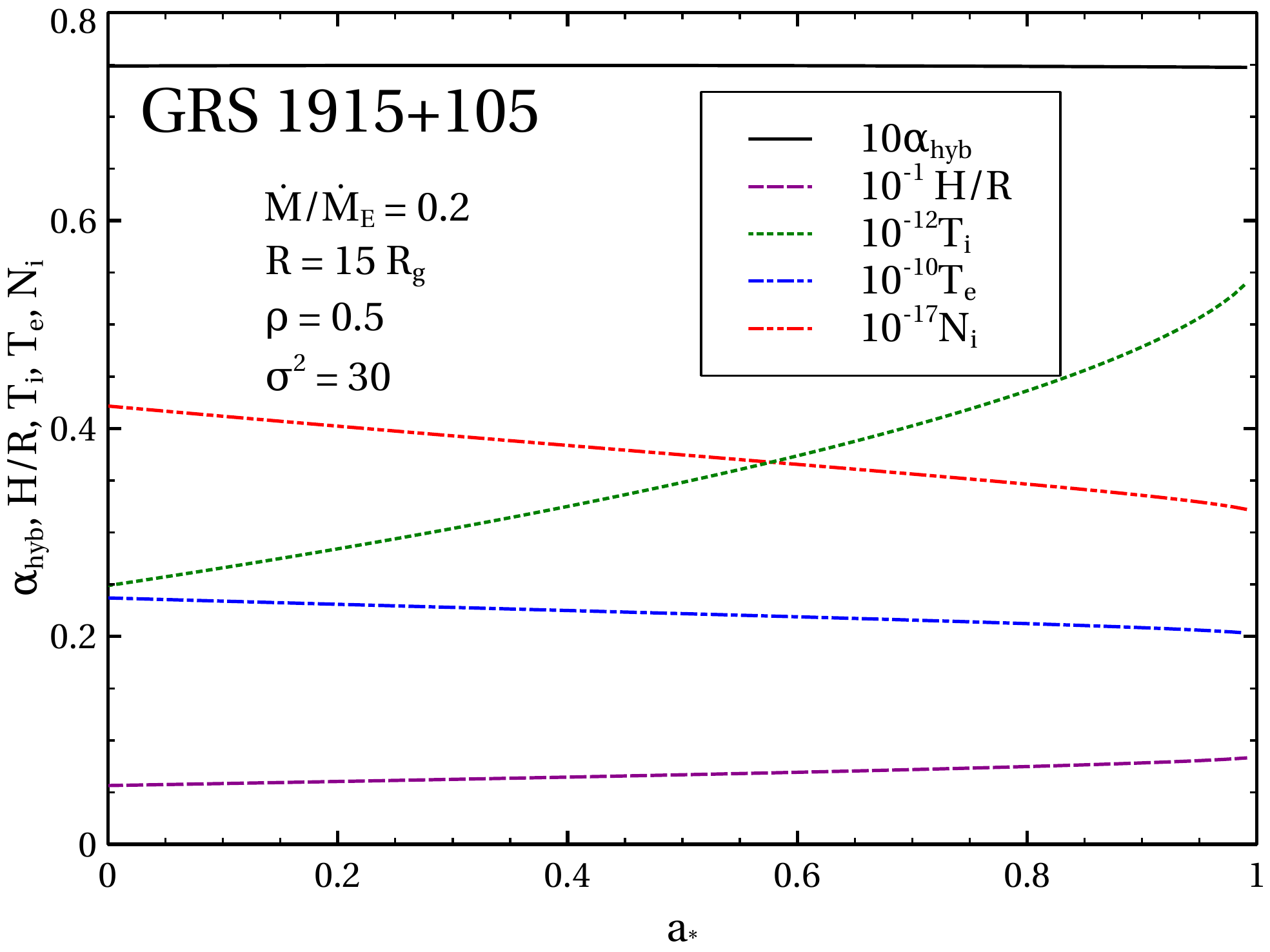}
	\caption{This figure shows how the quantities $10 \alpha_{\rm hyb}$, $10^{-1}H/R$, $10^{-12}T_{i}$, $10^{-10}T_{e}$ and $10^{-17}N_{i}$ (evaluated at $R = 15 R_{\rm g}$) vary with the black hole spin parameter $a_{*}$. The black hole mass is taken to be representative of GRS 1915+105, the accretion rate is $\dot{M}/\dot{M}_{E}=0.2$ and $\rho=0.5$, $\sigma^{2}=30$.}
	\label{fig:diskvsspin}
\end{figure}

Although our fiducial models assume a non-rotating black hole, the source GRS 1915+105 is thought to harbour a rotating black hole by some authors \citep{blum2009,miller2013}. It is therefore worth examining how the disc solutions for this source change with the black hole spin parameter $a_{*}$. Figure~(\ref{fig:diskvsspin}) shows $\alpha_{\rm hyb}$, $H/R$, the proton and electron temperatures and the density as a function of the black hole spin $a_{*}$. These quantities are evaluated at a radius of 15 $R_{\rm g}$. The linestyles are identical to those used in the lower panel of Figure~(\ref{fig:diskpara}). We note that $H/R$ and $\alpha_{\rm hyb}$ are very insensitive to $a_{*}$. When $a_{*}$ is increased from 0 to 0.98, the proton temperature increases by a factor $\approx 2$, while the electron temperature decreases by $\approx 20$\%. We will investigate the dependence of the infall timescale on $a_{*}$ later on in the paper.

\section{Inner disc collapse timescale and connection with X-ray timescales}
The radial infall timescale in an accretion disc is given by
\begin{equation}
t_{\rm infall} \equiv \frac{R}{v_{\rm R}} = \frac{R^2}{\nu} \, ,    \label{eq:k}
\end{equation}  
where $v_{\rm R}$ is the radial infall velocity (which depends upon the operative viscosity), $\nu \,\, ({\rm cm^2 s^{-1}}) \equiv \eta / N m = \alpha c_{s} H$ is the kinematic viscosity and $c_{s}$ is the sound speed. Using the disc model of \S~\ref{dm} the expression for the infall timescale becomes 
\begin{equation}
t_{\rm infall} = 492.5 \,\, \bigg( \frac{\dot{M}}{\dot{M}_{E}} \bigg)^{-1} M_{8} f_{1}^{1/2} f_{2}^{-1} \tau_{\rm es} R_{*}^2    \label{eq:l}
\end{equation}

We emphasize that Eq~(\ref{eq:l}) represents the infall timescale in the hot inner accretion disc and not in the cold, outer disc. Clearly, the timescale given by Eq~(\ref{eq:l}) increases with radius. In comparing with the observed X-ray timescales, we use the infall timescale evaluated at the outer edge ($R_{\rm out}$) of the hot, inner disc, and assume that this is the timescale over which the inner disc collapses. 

\begin{figure}
	\includegraphics[width=\columnwidth]{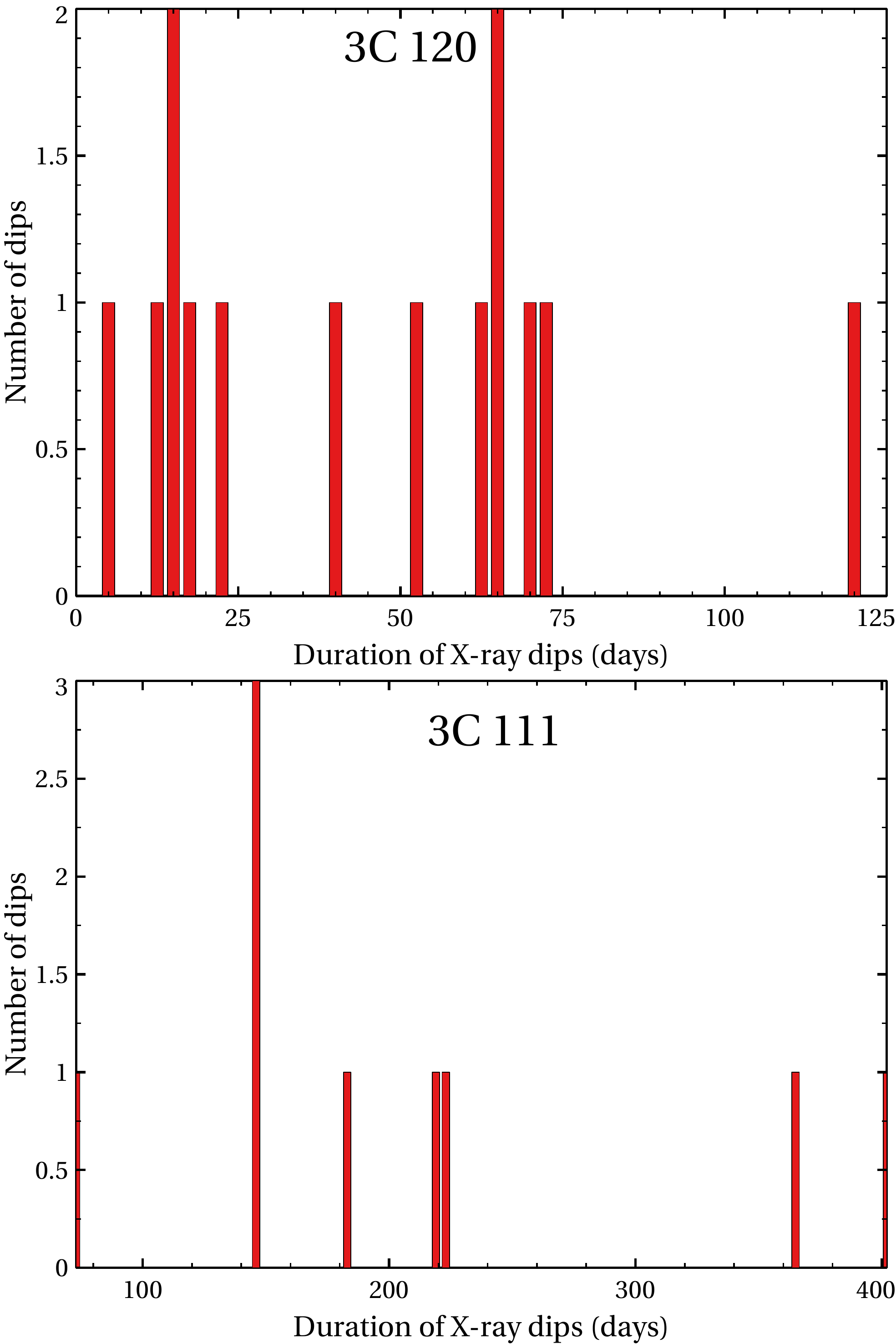}
	\caption{Histograms of the X-ray dip durations observed in 3C 120 (upper panel), and 3C 111 (lower panel)}
	\label{fig:hist}
\end{figure}

Figure~(\ref{fig:hist}) shows the distribution of observed X-ray dip timescales in 3C 120 \citep{chatterjee2009} in the upper panel, and 3C 111 \citep{Chatterjee2011} in the lower panel. We note that the dip durations in 3C 120 range from 5--120 days, with most of them in the range 5--75 days. Similarly, X-ray dip timescales for 3C 111 range from 73--402 days. \cite{belloni1997}, and \cite{yadav1999} report the observations of X-ray bursts in the galactic microquasar GRS 1915+105 with rise times ranging from 1--6 seconds. 
The X-ray spectrum in the quiescent (i.e., low intensity) states is hard, while it becomes soft during the burst (i.e., high intensity) states. Since the hot, inner disc is generally thought to be the source of hard X-rays, these observations suggest that the inner accretion disc collapses over the rise time of the burst. This idea is similar to that presented in \cite{yadav1999}, who compare the rise timescale with the viscous timescale of the Comptonized, sub-Keplerian halo ($t_{\rm vis}^{\rm h}$). Our motivation in calculating the infall timescale of the hot inner disc (Eq~\ref{eq:l}) is similar, with one crucial difference - instead of assuming a value for the Shakura-Sunyaev $\alpha$ parameter, it is self-consistently calculated using the viscosity mechanism outlined in \S~\ref{visc}.

Our aim in this paper is to compute the infall timescale given by Eq.~(\ref{eq:l}) and compare it with the observed X-ray dip timescales in 3C 111 and 3C 120 and the fast rise timescales in the X-ray bursting state for GRS 1915+105. In all cases, we have assumed a non-rotating black hole; this determines the relativistic correction functions $f_{1}$, $f_{2}$ and $f_{3}$ used in the disc structure equations. We have taken the value of Compton $y$ parameter to be unity and $\ln{\Lambda}=15$.  The black hole mass for 3C 120 is taken to be $M_{8} = 0.55$ \citep{peterson2004}, $M_{8} = 1.8$ for 3C 111  \citep{Chatterjee2011}, and  $M_8 = 1.24 \times 10^{-7}$ for GRS 1915+105  \citep{reid2014}. The values of the parameters $a_{\perp}$, $N_{\perp}$, $N_{\parallel}$, $\rho_{\parallel}$,  and $\gamma$ are given in Table 1 of \cite{candia2004} for different kinds of turbulence. For the sake of concreteness, we use the values corresponding to Kolmogorov turbulence. We also used the other turbulence models (i.e. Kraichnan and Bykov--Toptygin) and found that our results are quite insensitive to the specific turbulence model used. The accretion rates are taken to be $\dot{M}/\dot{M}_{E} = 0.3$ for 3C 120 \citep{chatterjee2009}, $\dot{M}/\dot{M}_{E} = 0.02 \pm 0.01$ for 3C 111 \citep{Chatterjee2011}, and $\dot{M}/\dot{M}_{E} = 0.2$ for GRS 1915+105 \citep{zdziarski2016}. The free parameters we are left with are $\rho$ and $\sigma^2$, which characterize the turbulence (\S~\ref{visc}) and  $R_{\rm out}$, which gives the extent of the hot, inner accretion disc. We will determine the parameter space (spanned by $\rho$, $\sigma^{2}$ and $R_{\rm out}$) that yield $t_{\rm infall}$ that match the observed ones.

\section{Results}

\subsection{Model parameters}
Our aim is to identify the combinations of free parameters that yield models with infall timescales that match the observed X-ray dip timescales. We run a grid of models with various combinations of the free parameters; the turbulence level ($\sigma^{2}$), the rigidity ($\rho$) and the outer radius of the corona ($R_{\rm out}$). Each of the models is required to satisfy the self-consistency conditions listed in \S~\ref{sc}.

\begin{figure}
	\includegraphics[width=\columnwidth]{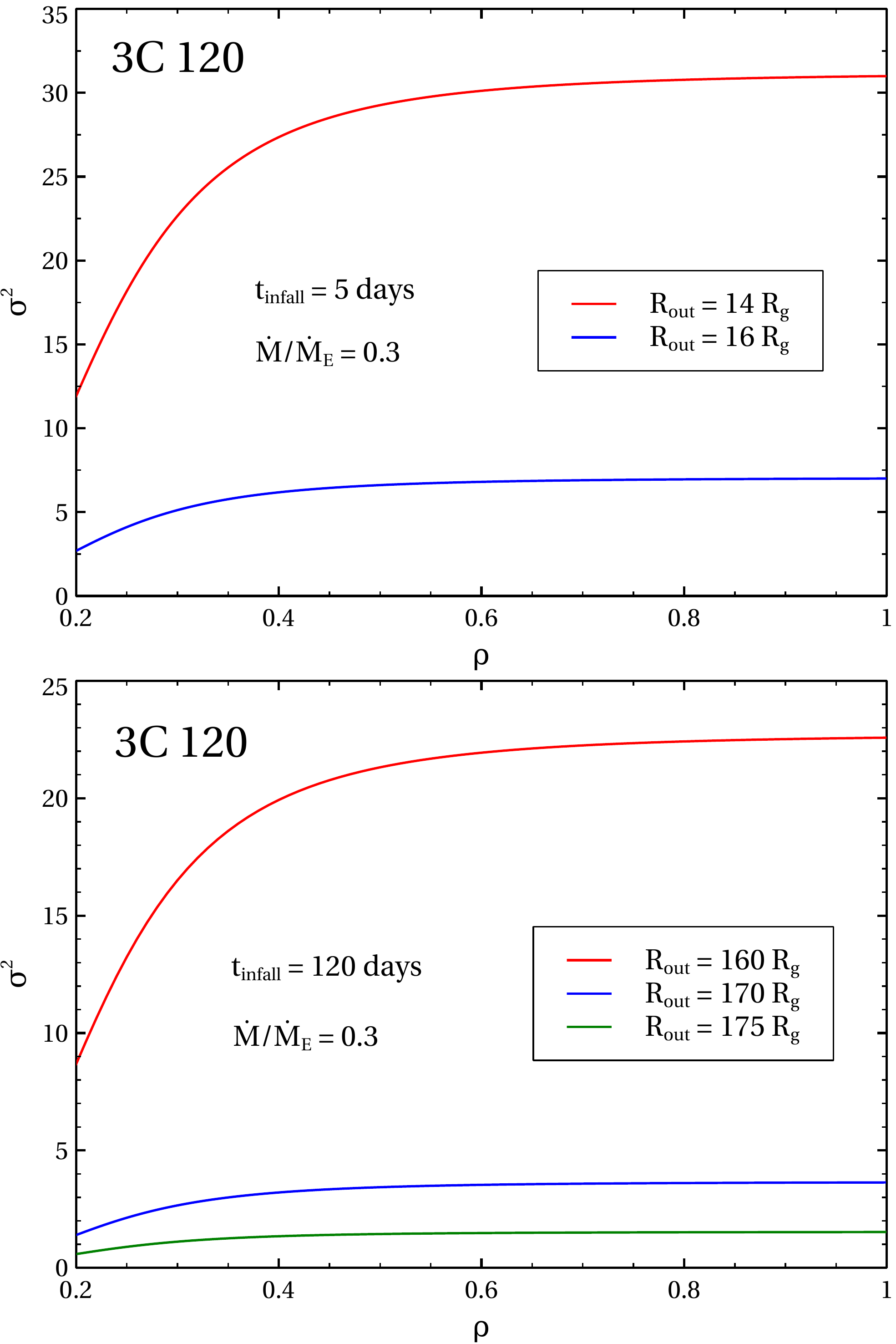}
	\caption{Upper panel: Parameter space corresponding to X-ray dip of 5 days in 3C 120, Lower panel: Parameter space corresponding to X-ray dip of 120 days in 3C 120}
	\label{fig:5_120days3c120}
\end{figure}

\begin{figure}
	\includegraphics[width=\columnwidth]{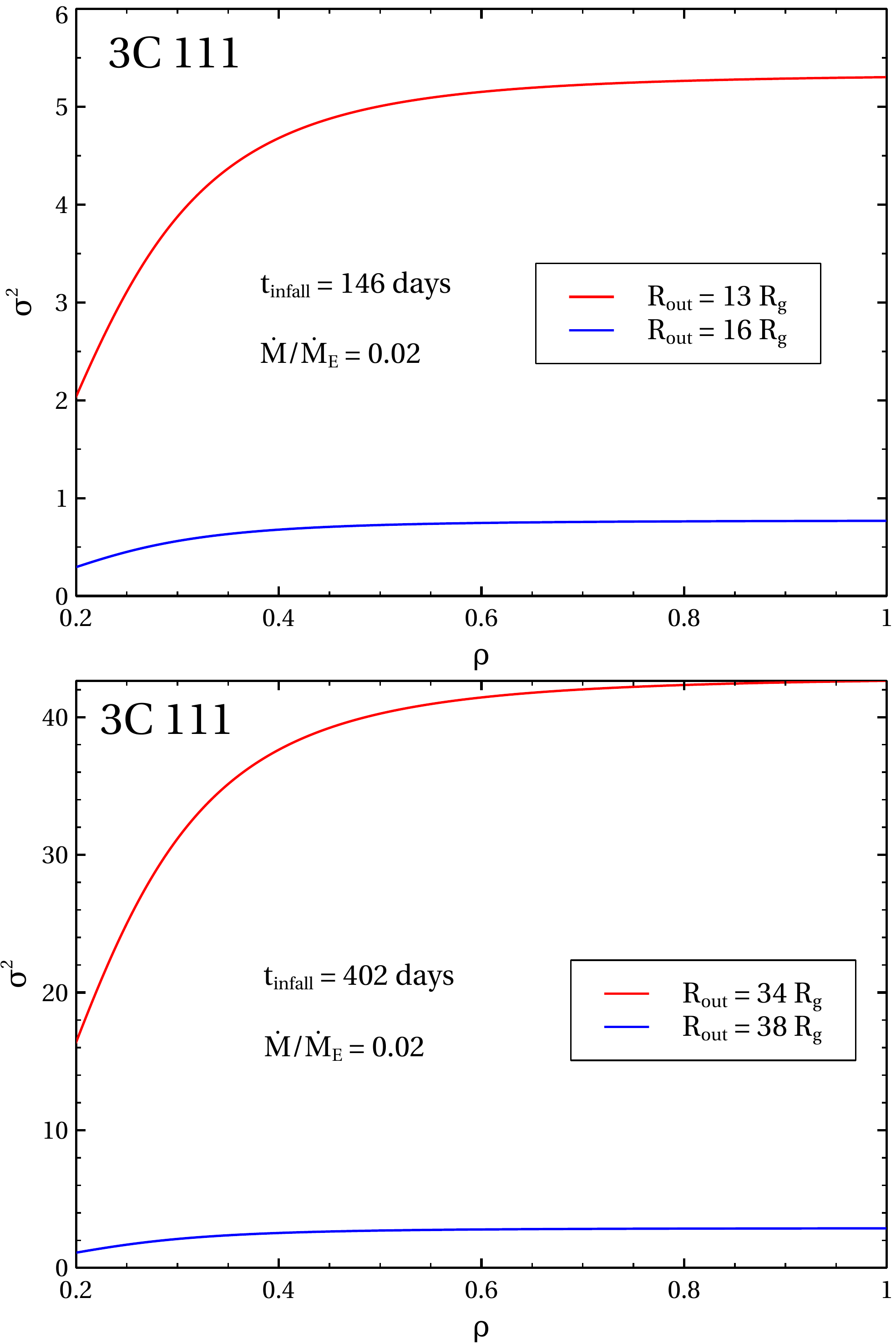}
	\caption{Upper panel: Parameter space corresponding to X-ray dip of 146 days in 3C 111, Lower panel: Parameter space corresponding to X-ray dip of 402 days in 3C 111}
	\label{fig:146_402days3c111}
\end{figure}

\begin{figure}
	\includegraphics[width=\columnwidth]{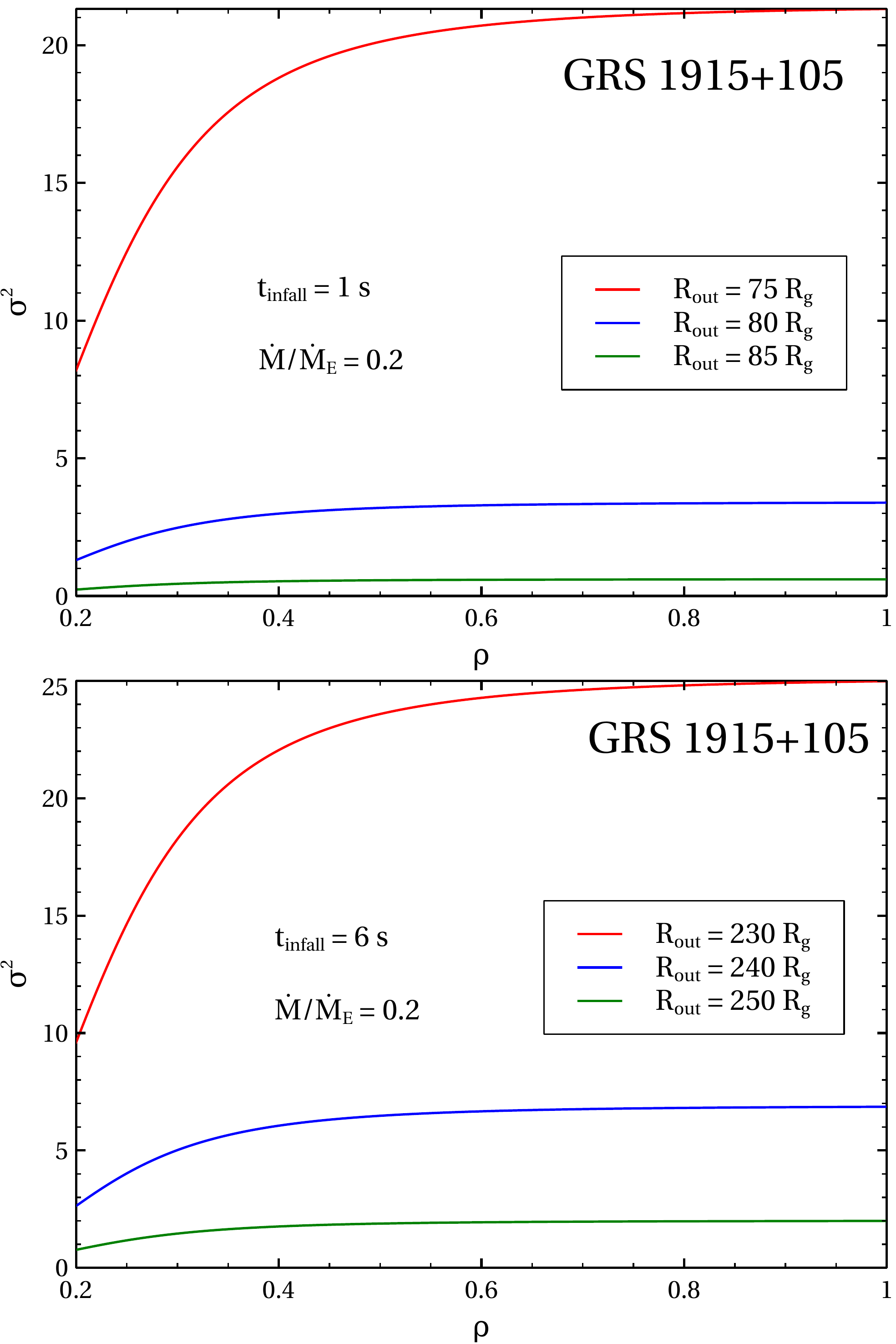}
	\caption{Upper panel: Parameter space corresponding to X-ray dip of 1 s in GRS 1915+105, Lower panel: Parameter space corresponding to X-ray dip of 6 s in GRS 1915+105}
	\label{fig:1_6sgrs}
\end{figure}

As mentioned earlier, the X-ray dip timescales for 3C 120 range from 5 to 120 days. The upper panel of figure~(\ref{fig:5_120days3c120}) depicts the combination of parameters that yield $t_{\rm infall}$ = 5 days while the lower panel shows the parameter combinations for $t_{\rm infall}$ = 120 days. Taken together, these plots reveal that the range of X-ray dip timescales for 3C 120 can be attained with our model for $0.2 \lesssim \rho < 1$, $ 1 \lesssim \sigma^{2} \lesssim 30$ (which corresponds to $0.04 < \alpha_{\rm hyb} < 0.07$) and $14 R_{\rm g} \lesssim R_{\rm out} \lesssim 175 R_{\rm g}$.
In figure~(\ref{fig:146_402days3c111}) we show the range of parameters that yield $t_{\rm infall}$ ranging from 146 to 402 days for 3C 111. It reveals that $146 < t_{\rm infall} < 402$ days can be attained for 3C 111 with $0.2 \lesssim \rho < 1$, $ 1 \lesssim \sigma^{2} \lesssim 40$ (which corresponds to $0.02 < \alpha_{\rm hyb} < 0.08$) and $13 R_{\rm g} \lesssim R_{\rm out} \lesssim 38 R_{\rm g}$. Figure~(\ref{fig:1_6sgrs}) is a similar plot which depicts the range of parameters that result in values of $t_{\rm infall}$ ranging from 1 to 6 seconds for the galactic microquasar GRS 1915+105. We see that the parameter range $0.2 \lesssim \rho < 1$, $ 1 \lesssim \sigma^{2} \lesssim 25$ (which corresponds to $0.03 < \alpha_{\rm hyb} < 0.07$) and $75 R_{\rm g} \lesssim R_{\rm out} \lesssim 250 R_{\rm g}$ is acceptable in this case. 

\begin{figure}
	\includegraphics[width=\columnwidth]{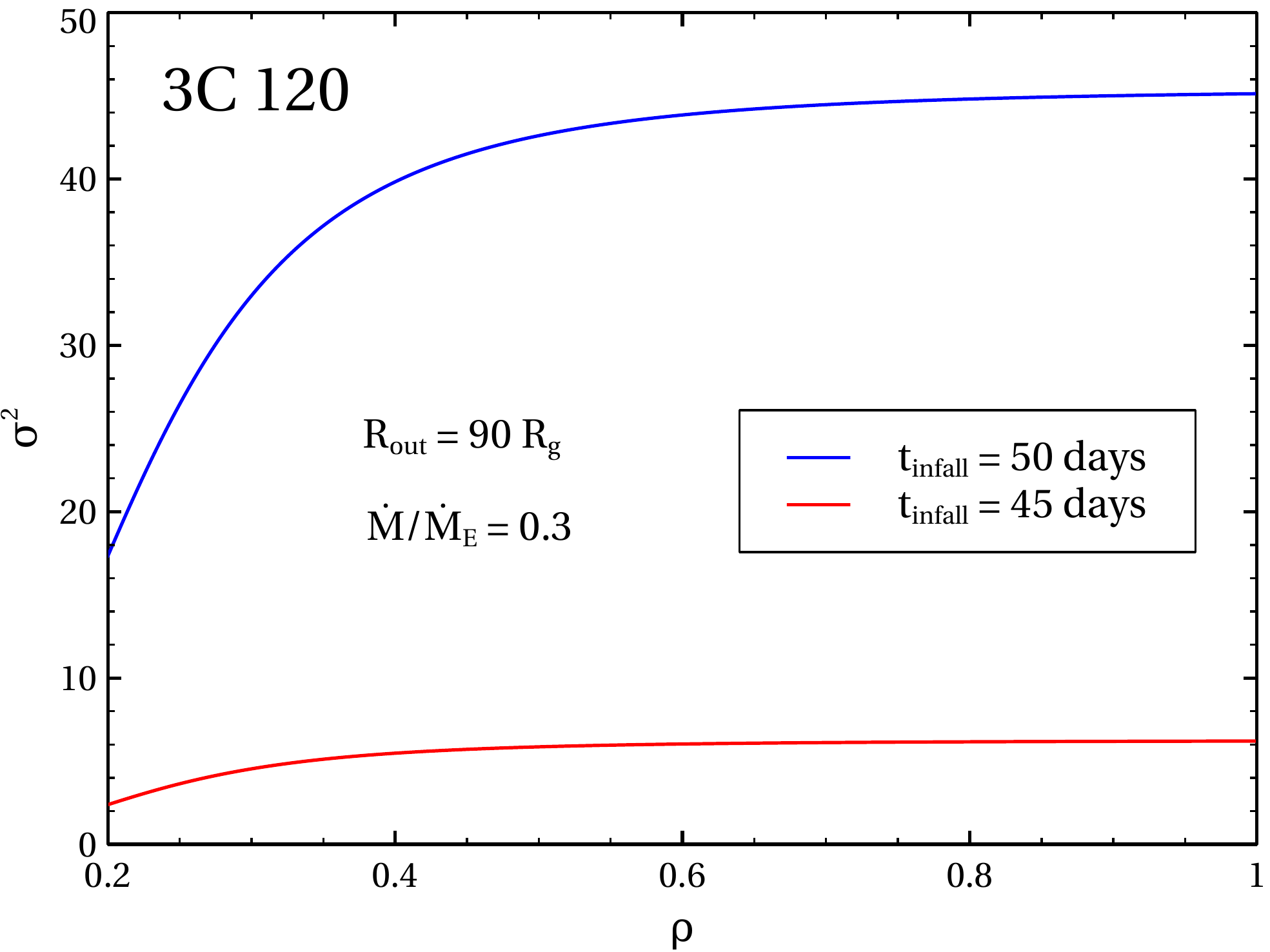}
	\caption{Parameter space corresponding to X-ray dip in the range 45--50 days in 3C 120}
	\label{fig:tvisc3c120}
\end{figure}

Figure~(\ref{fig:tvisc3c120}) explores the parameter space with $R_{\rm out}$ held constant. We find that infall timescale $t_{\rm infall}$ decreases with increasing $\rho$, while it increases with increasing $\sigma^2$. For 3C 120, we find that $t_{\rm infall} \approx$ 39--54 days with $R_{\rm out}$ held constant at $90 \, R_{\rm g}$. Similarly for 3C 111, the range for $t_{\rm infall}$ with $R_{\rm out} = 30 \, R_{\rm g}$ is $\approx$ 263--378 days.

\subsection{Sensitivity of $t_{\rm infall}$ to model parameters} \label{sensitivity}

\begin{figure}
	\includegraphics[width=\columnwidth]{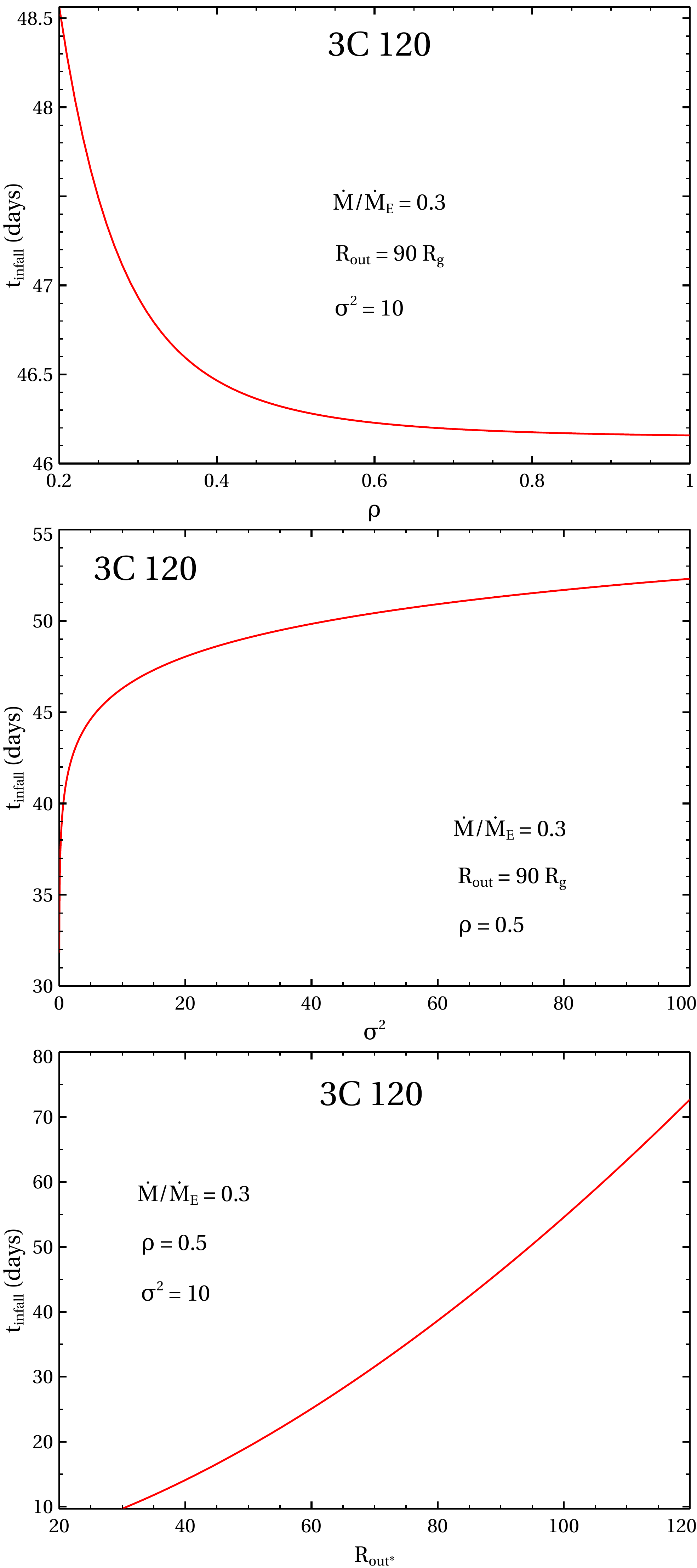}
	\caption{A depiction of how $t_{\rm infall}$ varies in response to changes in $\rho$, $\sigma^2$, and $R_{\rm out}$}
	\label{fig:sensitivity3c120}
\end{figure}

The first point evident from figures~(\ref{fig:5_120days3c120}), (\ref{fig:146_402days3c111}) and (\ref{fig:1_6sgrs}) is that the observed X-ray timescales for each source studied here can be explained with a reasonable parameter range in our model. The next obvious question to examine is the parameter(s) to which our results are most sensitive. To recapitulate, the quantity $\rho \equiv r_{L}/H$ is a measure of how ``tightly tied'' the protons are to the mean magnetic field. It is called the magnetic rigidity, and quantifies the extent to which the protons are magnetized. A low value for $\rho$ would indicate that the protons are strongly magnetized, while a high value (the upper limit being 1) would indicate the opposite. The quantity $\sigma^{2} \equiv \langle B_{r}^{2} \rangle /\langle B_{0}^{2} \rangle$ is the ratio of the energy density of the turbulent magnetic fluctuations to that of the ordered magnetic field, and can be regarded as a measure of the strength of the turbulence. The quantity $R_{\rm out}$ denotes the outer radius of the hot, inner accretion disc.
In order to examine the sensitivity of the result for $t_{\rm infall}$, we vary each parameter in turn
(while holding the rest of them fixed). The results of such a sensitivity analysis for 3C 120 are shown in table (\ref{tab:sensitivity}) and figure (\ref{fig:sensitivity3c120}). These results are applicable to the other two sources as well. The reference values of parameters for this study are: $\rho_{\rm ref} = 0.5$, $\sigma^{2}_{\rm ref} =10$, $R_{\rm out_{ref}} = 90 \, R_{\rm g}$. It is evident from figure (\ref{fig:sensitivity3c120}) that $t_{\rm infall}$ is a monotonically decreasing function of $\rho$, while it increases monotonically with $\sigma^{2}$ and $R_{\rm out}$.  In each row of table (\ref{tab:sensitivity}), the parameter shown in red is varied while the rest are held fixed. This table shows that the values for $t_{\rm infall}$ are least sensitive to variations in $\rho$ and most sensitive to variations in $R_{\rm out}$. Varying $R_{\rm out}$ by $\pm$ 10\% results in a $\approx \pm$ 15 \% variation in $t_{\rm infall}$. By contrast, varying $\rho$ by $\pm$ 10\% results in only a $\mp$ 0.1\% change in $t_{\rm infall}$, and a change of $\pm$ 10\% in $\sigma^2$ yields a $\pm$ 0.5\% change in $t_{\rm infall}$. As mentioned earlier, our fiducial model for GRS 1915+105 assumes a non-rotating black hole, while there is some evidence that it harbours a rotating black hole. In our fiducial model for GRS 1915+105, we find that increasing the black hole spin parameter $a_{*}$ from 0 to 0.98 necessitates a 12\% increase in $R_{\rm out}$ in order to yield an infall timescale $t_{\rm infall} = 1$ second. Conversely, if $R_{\rm out}$ is kept fixed at 75 $R_{\rm g}$, increasing $a_{*}$ from 0 to 0.98 results in a 17\% decrease in $t_{\rm infall}$.

\begin{table}
	\caption{Sensitivity analysis of parameters: fiducial model for 3C120}
	\label{tab:sensitivity}
	\begin{tabular}{|c|c|c| }

		\hline
		Parameter & \% change in reference & \% change in $t_{\rm infall}$\\[2pt] 
		\hline
		\textcolor{red}{$\rho$} & +10 & -0.1\\[2pt]
		($\rho_{\rm ref} = 0.5$, $\sigma^{2} = 10$, & -10 & +0.1\\[2pt]
		$R_{\rm out} = 90 \, R_{\rm g}$) & & \\[2pt]
		\hline
		\textcolor{red}{$\sigma^2$} & +10 & +0.5\\[2pt]
		($\sigma^{2}_{\rm ref} = 10$, $\rho = 0.5$, &-10 & -0.5 \\[2pt]
		$R_{\rm out} = 90 \, R_{\rm g}$) & & \\[2pt]
		\hline
		\textcolor{red}{$R_{\rm out}$} & +10 & +15\\[2pt]
		($R_{\rm out_{ref}} = 90 \, R_{\rm g}$, & -10 & -15\\[2pt]
		$\rho = 0.5$, $\sigma^{2} = 10$) & & \\[2pt]
		\hline
		
	\end{tabular}
\end{table}

\subsection{Summary and Conclusions} \label{conclusions}

To recapitulate, X-ray dips in systems such as AGN 3C 120 and 3C 111, and the rise time of the bursts in X-ray intensity in galactic microquasar GRS 1915+105 are thought to arise from the collapse of the inner accretion disc, which occurs over the radial infall timescale (Eq.~\ref{eq:l}). 
X-ray dip timescales for 3C 120 are in the range 5--120 days with most dips happening over 5--75 days. The dip timescales for 3C 111 are in the range 73--402 days. In the galactic microquasar GRS 1915+105, the rise time of X-ray bursts (which can be thought of as the dip timescale for hard X-ray intensity) is $\approx$1--6 seconds.
The radial infall timescale of the hot inner accretion disc is crucially governed by the operative viscosity, which is commonly parametrized via the Shakura-Sunyaev $\alpha$ parameter ($\nu = \alpha \, c_{s} \, H$). Save for the constraint $0 < \alpha < 1$, not much is known about it, and it's generally treated as a fitting parameter. In this work, we have evaluated the coefficient of viscosity operative in the hot, inner regions of the accretion discs. The plasma in these regions is so hot that the protons are collisionless; they rarely collide with each other, but they can be scattered via interactions with turbulent magnetic fields. There is a considerable body of literature that gives the diffusion coefficient derived from detailed simulations of cosmic ray protons diffusing across a large-scale magnetic field in the presence of turbulence. The mean free path extracted from this diffusion coefficient is used to compute the operative viscosity, which we term the ``hybrid'' viscosity (since it is neither due to proton-proton collisions nor due to magnetic field stresses, but due to protons bouncing off magnetic scattering centres). This approach is similar to that used to construct a cosmic ray viscosity \citep{earl1988}. Computing the hybrid viscosity enables us to construct accretion disc models (which incorporate a physically motivated viscosity) and calculate the collapse timescale for the inner disc.

Our results have three free parameters: the ratio of energy density in the turbulent magnetic field to that in the large-scale magnetic field (the turbulence level $\sigma^{2}$), the ratio of the proton Larmor radius to the disc height (the proton rigidity $\rho$) and the outer radius of the hot, inner accretion disc--corona $R_{\rm out}$. Different combinations of these parameters yield radial infall timescales (Eq.~\ref{eq:l}) that can match the ones observed for 3C 120, 3C 111, and GRS 1915+105. Figures (\ref{fig:5_120days3c120})--(\ref{fig:1_6sgrs}) show the space spanned by these three parameters and gives a comprehensive idea of the scope of our model. Each point in these figures represents an accretion disc model such as the one shown in figure~(\ref{fig:diskpara}). We find that the infall timescale $t_{\rm infall}$ is not very sensitive to variations in the parameters $\rho$ and $\sigma^{2}$. It is very insensitive to $\rho$ (a 10\% change in $\rho$ results in only a 0.1\% change in $t_{\rm infall}$) and quite insensitive to $\sigma^{2}$ as well (a 10\% change in $\sigma^{2}$ results in a  0.5\% change in $t_{\rm infall}$). By contrast, a 10\% change in the outer radius of the hot inner disc ($R_{\rm out}$) results in a 15\% change in $t_{\rm infall}$ - our results are thus most sensitive to $R_{\rm out}$. To summarize,
\begin{itemize}
\item
The collapse of the inner accretion disc occurs over the radial infall timescale, which is governed by the operative viscosity. Instead of relying on values assigned to the dimensionless $\alpha$ viscosity parameter, we have outlined a prescription for the viscosity operative in the hot, inner part of black hole accretion discs and constructed simplified disc models using this prescription. The disc infall timescales ($t_{\rm infall}$) obtained with this model are in good agreement with X-ray observations of 3C 120, 3C 111 and GRS 1915+105. Together with models such as \cite{shende2019}, our work here outlines a plausible scenario for episodes of (inner) disc collapse accompanied by blob ejection.

\item
We have found that our results for $t_{\rm infall}$ are fairly insensitive to the parameters $\sigma^{2}$ and $\rho$ that are used to compute the coefficient of hybrid viscosity. The values of the Shakura-Sunyaev $\alpha$ parameter arising from our viscosity prescription range from 0.02 to 0.08.  
\item
The model predictions for the inner disc infall timescale are most sensitive to the disc outer radius $R_{\rm out}$, with larger $R_{\rm out}$ yielding larger values for $t_{\rm infall}$. For 3C 120, we find that our models require $14 \lesssim R_{\rm out} \lesssim 175\,R_{\rm g}$ to match the observational results for X-ray dip timescales of 5--120 days. For 3C 111, we require $13 \lesssim R_{\rm out} \lesssim 38\,R_{\rm g}$ for dip timescales of 73--402 days and for GRS 1915+105 we require $75 \lesssim R_{\rm out} \lesssim 250\,R_{\rm g}$ for dip timescales of 1--6 seconds.

\item
These values of $R_{\rm out}$ might seem somewhat large in comparison with the values of $\approx$ 20 $R_{\rm g}$ quoted by \cite{reis2013} for the size of hot, X-ray emitting coronae. For GRS 1915+105, for instance, we find that $75 \lesssim R_{\rm out} \lesssim 250\,R_{\rm g}$. However, the accretion rates used in \cite{reis2013} are around 0.01 times the Eddington value. On the other hand, we use $\dot{M}/\dot{M}_{E}=0.2$ for GRS 1915+102, following \citep{zdziarski2016}. If we were to use $\dot{M}/\dot{M}_{E}=0.02$, our model requires $12 \lesssim R_{\rm out} \lesssim 22\,R_{\rm g}$ for dip timescale of 1 second. We also note that the size of the hot, post-shock region in GRS 1915+102 has been estimated to be 90--132 $R_{\rm g}$ \citep{nandi2001}.
\end{itemize}

\section{Acknowledgements}

MBS acknowledges a PhD student fellowship from IISER, Pune. PS acknowledges helpful conversations with Ritaban Chatterjee. We acknowledge critical comments from anonymous referees that have helped us greatly in improving this paper.





\bibliographystyle{mnras}
\bibliography{draftfinal6}



\appendix
\section{Two-Temperature Accretion Disc, Comptonized Model} \label{append}

In a cylindrically symmetric accretion disc with no vertical structure, the relevant component of the stress arising from the hybrid viscosity is given by (SBK96)
\begin{equation}
\alpha_{\rm hyb} P \equiv -\eta_{\rm hyb} R \frac{d \Omega_{\rm kepl}}{dR}   \label{eq:aa}
\end{equation}

The basic disc structure equations neglecting radiation pressure are same as used in \cite{eilek1983} and are given by

\begin{equation}
P=\frac{GMm_{\rm p} N_{i} H^2 f_1}{R^3}\, \,\,({\rm Vertical\,hydrostatic\,equilibrium})   \label{eq:ab}
\end{equation}

\begin{equation}
\alpha P = \frac{(GMR)^{1/2} \dot{M} f_2}{4 \pi R^2 H}\,\,\,({\rm Radial \, momentum \, transport})   \label{eq:ac} 
\end{equation}

\begin{equation}
\frac{3}{8\pi} \frac{G M \dot{M}}{R^3 H} f_3 = 3.75 \times 10^{21} m_{\rm p} \ln{\Lambda} N_i^2 k_{\rm B} \frac{(T_i - T_e)}{T_e^{3/2}}\,\,\,({\rm Energy\,equation}) \label{eq:ad}
\end{equation}

\begin{equation}
P = N_{i} k_{\rm B} (T_{i} + T_{e}) \,\,\,({\rm Equation\,of\,state})   \label{eq:ae}
\end{equation}

\begin{equation}
T_{e} = \frac{m_{e} c^2 y}{4 k_{\rm B}} \frac{1}{\tau_{\rm es} g(\tau_{\rm es})}\,\,\,({\rm Comptonized \, electrons}) \label{eq:af}
\end{equation}

\begin{equation}
\tau_{\rm es} = N_i \sigma_{\rm T} H \,\,\,({\rm Definition \,of \,optical \,depth})    \label{eq:ag}
\end{equation}

where $y$ is the Compton y-parameter, $\sigma_{\rm T}$ is a Thomson scattering cross section, $\tau_{\rm es}$ is the electron scattering optical depth and the function $g(\tau_{\rm es}) \equiv 1 + \tau_{\rm es}$. We have taken the value of Coulomb logarithm $\ln{\Lambda}$ to be 15 in our calculations. In deriving these equations, we have assumed the steady state conditions. Also, we assume that the disc is quasi-Keplerian, so that the azimuthal velocity $v_{\phi}$ is essentially the Keplerian velocity $= \sqrt{GM/R}$, and the radial drift velocity $v_{R}\ll v_{\phi}$. Since there is no vertical structure by assumption, we can adopt the scaling relations $\partial P /\partial z \sim -P/H$, and $z \sim H$. Eq.(\ref{eq:ab}) denotes the verical hydrostatic equilibrium, whereas Eq.(\ref{eq:ac}) represents the angular momentum transport. Ion thermal balance is given by Eq.(\ref{eq:ad}), whereas unsaturated Compton cooling is given by Eq.(\ref{eq:af}). The system of equations is completed with the equation of state, Eq.(\ref{eq:ae}), and the optical depth definition, Eq.(\ref{eq:ag}). The factors $f_1$, $f_2$ and $f_3$ are the relativistic correction factors for the metric under consideration \citep{novikov1973,page1974}. For our calculations, we have used the values of $f_1$, $f_2$ and $f_3$ for a Schwarzschild black hole ($a/M = 0$). 

Assuming $T_i \gg T_e$ and making the definitions $M_8 \equiv M / 10^8 M_{\odot}$, and $R_{*} \equiv R/(G M /c^2)$, equations (\ref{eq:ab})--(\ref{eq:ag}) yield the following analytical solutions: 

\begin{equation}
T_{i} = 1.08 \times 10^{13} \frac{\dot{M}}{\dot{M}_{E}} f_{2} \tau_{\rm es}^{-1} \alpha^{-1} R_{*}^{-3/2}      \label{eq:ah}
\end{equation}

\begin{equation}
T_{e} = 1.48 \times 10^{9} y \tau_{\rm es}^{-1} [g(\tau_{\rm es})]^{-1}   \label{eq:ai}
\end{equation}

\begin{equation}
N_{i} = 1.02 \times 10^{11} \bigg(\frac{\dot{M}}{\dot{M}_{E}} \bigg)^{-1/2} M_{8}^{-1} f_{1}^{1/2} f_{2}^{-1/2} \alpha^{1/2} \tau_{\rm es}^{3/2} R_{*}^{-3/4}       \label{eq:aj}
\end{equation}
where $\dot{M}/\dot{M}_{E}$ represents the accretion rate in units of Eddington rate ($\dot{M}_{E} \equiv L_{\rm E}/c^2$, where $L_{\rm E} \equiv 4 \pi G M m_{\rm p} c /\sigma_{\rm T}$ is the Eddington luminosity and $\sigma_{\rm T}$ is the Thomson electron scattering cross section).
It may be emphasized that $\alpha$ is a free parameter in the above solutions.


\label{lastpage}
\end{document}